# From RE-211 to RE-123. How to control the final microstructure of superconducting single-domains

Cloots R.[1,*], Koutzarova T.[1], Mathieu J.-P.[1] and Ausloos M.[2,]

[1] SUPRATECS, Chemistry Institute B6, University of Liège, Sart Tilman, B-4000 Liège, Belgium
[2] SUPRATECS, Physics Institute B5, University of Liège, Sart Tilman, B-4000 Liège, Belgium
* e-mail: rcloots@ulg.ac.be
e-mail: Marcel.Ausloos@ulg.ac.be

ABSTRACT:

This paper reviews the usual techniques for producing YBCO-type single-domains and the microstructure of the as-obtained samples. The problems of seed dissolution and parasite nucleations are discussed in details. Formation of microstructural defects, such as pores and cracks, are examined. An important part of this review is devoted to the study of the influence of RE-211 particles [$RE_2BaCuO_5$ where RE denotes Y, Yb, Nd, Sm, Dy, Gd, Eu or a mixture of them. Generally $Nd_4Ba_2Cu_2O_{10}$ is preferred to $Nd_2BaCuO_5$] on the microstructure and properties of RE-Ba-Cu-O single-domains. Trapping/Pushing theory is described in order to explain the spatial distribution of RE-211 particles in the RE-123 [$(RE)Ba_2Cu_3O_{7-\delta}$] monoliths. Formation of RE-211-free regions is discussed. Different ways to limit the RE-211 coarsening are reviewed. Microstructural defects in the RE-123 matrix caused by the RE-211 particles are presented. It is also shown that RE-211 particles play a significant role on the mechanical properties of single-domain samples. We finish this review by discussing the Infiltration and Growth process as a good technique to control the microstructure.

## I. INTRODUCTION

During the past fifteen years, enormous efforts have been made in the development of high-$T_c$ oxide superconductors that are available for practical applications. The high-temperature (Rare Earth)-Ba-Cu-O [(RE)BCO] single-domain bulk superconductors are attractive for various engineering applications, *e.g.* as in magnetic bearing (flywheel, ) [1-

4] or as trapped flux magnets (magnetic separation, ) [5,6]. They have the potential to yield high critical current density, $J_c$, even under high applied magnetic fields. Moreover, the magnetic field which can be trapped in (RE)BCO single-domains can be more than twice larger than in conventional permanent magnets [5,6].

The number of applications involving polycrystalline RE-123 samples is limited by the weak-link nature of grain boundaries which results in low $J_c$ [7,8]. The weak link properties are due to secondary non-superconducting phases, misorientations between grains, chemical and physical inhomogeneities at the grain boundary

Whether in trapped magnetic fields or in bearings, properties are strongly dependent on the size of the shielding current loop, the critical current density ($J_c$) and the grain microstructure [1,9,10]. Large superconducting single-grains, so-called single-domains, are thus required [11-16].

Due to the peritectic decomposition of the RE-123 phase:

$$(RE)\text{-}123 \_ (RE)\text{-}211 + liquid \quad (1)$$

large single-crystals cannot be grown by a slow cooling process from a liquid phase (see figure 1). Nevertheless, large single-grains can be produced by peritectic recombination from the RE-211°+°liquid state. In this case, quasi-crystals with a lot of defects can be obtained [11,17-19].

Defects such as the presence of the $RE_2BaCuO_5$ (RE-211) secondary phase in the RE-123 matrix, which results from the incomplete peritectic recombination, can act as pinning centres for vortices in the superconducting state and thus help to increase $J_c$ [20-24]. On the other hand, macro-cracks present in textured samples have been found to dramatically damage the superconducting properties [25-29]. Therefore, very careful preliminary investigations and fine control of the microstructure of (RE)BCO single-domains are key factors for developing successful practical use of these materials.

In this review, the attention is paid to the microstructure of single-domains by considering especially the role played by the RE-211 particles.

In section 2, the main different techniques to produce YBCO-type single-domains are reviewed. In the third section, the microstructure of the as-obtained samples is examined in details. The problems of seed dissolution and parasite nucleations are discussed. Formation of microstructural defects, such as pores and cracks, is exposed. The study of the influence of RE-211 particles on the microstructure and properties of (RE)BCO single-domains takes an important place in this section. Phenomena like pushing/trapping, RE-211-free regions, Oswald ripening are discussed. It is also shown in this third section that even if RE-211 particles cause the formation of defects, they also play a beneficent role on the mechanical properties of the single-domains.

In section 4, we present the Infiltration and Growth (IG) process. This process can be an alternative to the classical Top-Seeded Melt Textured Growth (TSMTG) process in order to control homogeneity of the RE-211-particles distribution in the RE-123 matrix single-domain. The other advantages of this technique are also presented.

## II. EXPERIMENT

The microstructure of single-domain (RE)BCO superconductors is highly dependent on the processing conditions and the presence or not of additives [19,27,30-36]. A variety of melt processing techniques has been developed to produce single-domain (RE)BCO superconductors [11,18,37-45]. All these techniques are based on the peritectic reaction:

$$2\ (RE)Ba_2Cu_3O_{7-\delta\ (s)} \_ (RE)BaCuO_{5\ (s)} + Ba_3Cu_5O_{8\ (l)} + y\ O_{2\ (g)} \quad (2)$$

This reaction occurs between 1000°C and 1100°C (the peritectic temperature, $T_p$) depending on the RE nature. In this process, the solid RE-123 phase decomposes at $T_p$ to form a solid RE-211 (or 422 for $Nd_4Ba_2Cu_2O_{10}$) phase and a Ba- and Cu-rich liquid phase according to the RE-Ba-Cu-O phase diagram (figure 1) [46].

The solid RE-123 phase is then re-formed under cooling the so-called RE-211 + liquid state to a temperature below $T_p$ [47], according to the following reaction:

$$RE\text{-}211_{(s)} + [3\ BaCuO_2 + 2\ CuO]_{(l)} \_ 2\ RE\text{-}123_{(s)} \quad (3)$$

In order to increase the kinetic of the process, the system is cooled down to a temperature about 30 to 40°C under $T_p$, creating thus a strong undercooling.

The partial dissolution of the RE-211 particles supplies the rare earth element necessary for the growth of the RE-123 phase. After processing down to room temperature, RE-211 particles are usually trapped in the RE-123 grains, due to an incomplete dissolution of RE-211 particles. These entrapped RE-211-particles in the RE-123 matrix after the process can be seen on figure 2.

As a consequence there is also a liquid phase segregation at the grain boundaries. Figure 3 shows an optical micrograph of a multigrain sample; liquid phase segregation can be observed at the grain boundary between two misoriented grains. Numerous pores are also visible. A heterogeneous microstructure is thus obtained. The RE-211 inclusions are effective pinning centres but also increase the toughness of the single-domain [48-50], while the segregation of liquid phase at the grain boundaries is a major factor limiting the obtention of good superconducting properties by creating weak links between adjacent superconducting grains [39,51]. It is thus desirable to add extra RE-211 particles in the starting composition (i) to prevent the loss of the liquid phase and (ii) to avoid the presence of liquid phase at the grain boundaries by shifting on the right the equation 2 [11,39]. Moreover, for good pinning properties it is very important to control the RE-211 particle size and their distribution in the RE-123 single-domain matrix [16,19,20,38,52-54].

However, melt-texturing growth processing leads to the formation of multi-grains due to the presence of several-nucleation centres (see figure 4). The grains are separated by grain boundaries, called natural grain boundaries and have misorientation angles of more than 10° [17,18,55].

The seeding technique was proposed to overcome the problem of the multi-nucleation process. The top-seeded melt-textured growth process (TSMTG) [13,16,47,56,57] has thus been widely adopted to prepare large (RE)BCO single-domains, which trap high magnetic fields up to 3 T at 77 K [6,58]. This process is based on single-domain growth by epitaxy starting from a seed: mainly a small single crystal with a higher melting temperature compared to the (RE)BCO single-domain preforms, and with similar crystal structure.

Nd-123 or Sm-123 single-crystals (Figure 5) are usually used as seeds for producing (RE)BCO single-domains. They have a higher peritectic decomposition temperature (1080°C and 1100°C respectively) than other RE-123 and similar lattice parameters to other RE-123 [12,59-62]. MgO (001) single-crystal can also be used [57,63].

During the slow cooling process, just below the peritectic temperature $T_p$, the growth follows the orientation of the seed and proceeds at best throughout the entire volume of the precursor material. In this way, a single-grain material, a so-called single-domain, can be obtained (see figure 6). A single-domain is thus defined as a quasi single-crystal containing many imperfections or defects. In this technique the single-grain orientation is controlled by the crystallographic orientation of the seed. Thus, this process allows overcoming the problem of weak links due to the presence of grain boundaries.

There are two different ways to perform the cooling of the (RE)BCO single-domain, *i.e. (i)* in a thermal gradient or *(ii)* isothermally.

*(i)* A thermal gradient can be applied to control carefully the single-domain growth [15,47,59,64,65]. The area near the seed is always maintained at a lower temperature than the rest of the sample. Thereby, secondary nucleation around the growing single-grain is strongly limited. By using a cold finger localized just above the single-crystal seed at the top of the RE-123 matrix, it is possible to create in-situ a very good radial thermal gradient [15,58,66]. Large single-domains can be grown by this technique. Nevertheless, the microstructure of the as-obtained single-domains reveals the presence of a lot of defects, their concentration increasing going away from the initial position of the seed [15,64,67].

*(ii)* The second cooling technique gives rise to the growth of a single-domain from an isothermally undercooled melt [18,48]. This gives us the possibility to grow more than one particular sample per thermal cycle in one run [68]. It is thus possible to consider an extensive production of (RE)BCO single-domains. Nevertheless the size of the single-domain remains limited when compared to the (RE)BCO single-domain grown by the cold finger technique. Notice that the microstructure of the single-domain is more homogeneous when it has been grown by an isothermal process than with the thermal gradient process.

In order to increase the size of the single-grain, joining techniques were developed starting with small well oriented (RE)BCO domains [14,55,69-74]. Joining two single-domains leads to the formation of an artificial low-angle grain boundary, whence a high connectivity between domains is required in order to obtain a high intergranular (now, intragranular) current.

Various techniques for joining (RE)BCO single-domains have been developed: *(i)* a natural joining using a multi-seeding technique [37,57,73,75-77]; *(ii)* diffusion bonding of polished surface under uniaxial pressure in absence of soldering agents; and *(iii)* artificial joining by using lower peritectic temperature rare earth compounds like $YbBa_2Cu_3O_7$, $ErBa_2Cu_3O_7$ or $TmBa_2Cu_3O_7$ as soldering agents [55,69,71,72,74,78]. Silver has been also used to reduce locally the peritectic temperature of the RE-123 material [14,70,73].

## III. MICROSTRUCTURE

### 1. Overview of material microstructure

Top-Seeded Melt-Textured (TSMT) (RE)BCO single-domains generally grow in a parallelepipedic form with (100), (010) and (001) crystal habit planes (see figure 7). In general the seeded peritectic solidification can be mainly divided into three steps. *(i)* First, the growth starts following a *sympathetic nucleation* process from the single-crystal seed. The partially molten mixture of (RE-211 + liquid) is in contact with a (001) surface at the bottom of the seed and on the (100)/(010) surfaces at the border of the crystal seed. Thus, five different domains can be grown from the single-crystal seed. *(ii)* Then subsequently a *facet development* and *(iii)* a *continuous growth* process can take place. Thus RE-123 single-domains can grow continuously from the seed crystal over the whole volume of the sample [53].

Interestingly Endo *et al.* [79] demonstrated how to obtain a single-domain with rectangular shape in the *ac*-plane using the *ac*-plane of a Sm-123 single crystal as a seed. The single domain has a rectangular shape due to the fact that the growth rate in the (100) and (110) directions are faster than in the (001) direction.

The general growth morphology of the surface of a single domain can be observed by Scanning Electron Microscopy (SEM). The surface of the single-domain in the vicinity of the seed is characterized by the presence of parallel lines perpendicular to the macroscopic growth direction of the domain (See figure 8). Under higher magnification it was observed that these lines are composed of RE-123 layers with parallel edges and therefore correspond to the *a*- or *b*-axis facet lines derived from the tetragonal symmetry in the *a-b* plane of the single-domain [24]. This is because the RE-123 has a tetragonal crystal structure at the grain growth temperature. These lines are indicative of the crystallographic structure and are not supercurrent limiting grain boundaries [11]. The presence of continuous facet lines in this region results from a uniform grain growth in the vicinity of the seed and a homogeneous microstructure results. When the distance from the seed increases, the RE-123-layers are more and more fragmented and the continuous facet lines characteristic progressively disappears. Therefore, in these regions, the growth front is non-uniform leading to the presence of microstructural inhomogeneities [11,24].

The transition from a uniform to a non-uniform growth morphology may be understood from the peritectic solidification process. In the melt-texturing growth process,

RE-211 particles dissolve into the liquid phase to provide the $RE^{3+}$ cations required for the solidification of the RE-123-phase. Consequently, a homogeneous distribution of $RE^{3+}$ cations in the liquid is required to maintain a uniform growth front. When the $RE^{3+}$ distribution is non-uniform in the liquid phase, local (RE)BCO dendrites may be formed [11]. Thus any change in local density of RE-211 particle subsequently results in a local modification of the $RE^{3+}$ concentration. It results in a local growth inhomogeneity. According to a detailed analysis [24] of the distribution of Y-211 inclusions in the Y-123 matrix made by Lo *et al.* it was revealed that the volume fraction of Y-211 inclusions increases gradually from the seed, tending to a saturated value of about 30% at a distance of about 4 mm from the crystal seed. In this region, the transition occurs between a uniform to a non-uniform growth front morphology. Therefore, the local non-uniform distribution of $RE^{3+}$ cations may cause some microstructural inhomogeneity.

## 2. Influence of seed dissolution on the microstructure of (RE)BCO single domain

The presence of the seed on the top of the single domain preform obviously influences the microstructure of the as-grown single-domain. In this sub-section, attention is paid on the single-crystal to be used as a seed, the dissolution of the seed and the resulting microstructure.

In order to have an effective TSMTG process, the seed must *(i)* be chemically inert toward the precursor or melt during the melt-textured growth process, *(ii)* remain solid at the peritectic temperature of the RE-123 phase, and *(iii)* have similar crystallographic lattice parameters as compared to the RE-123 phase. According to these, Sm-123 and Nd-123 single-crystals seem to be the most suitable seeds for growing high quality (RE)BCO single-domains [12,59-62].

Although the processing temperature is lower than the melting point of the single crystal seed, partial or complete dissolution of the seed is often observed and thought to affect the growth mode of the single-domain [61,80,81]. The degree of seed dissolution depends on the processing variables such as the maximum processing temperature, the holding time, the seeding method (cold or hot seeding method), the nature of seeds and compact [61], ...

Three different growth modes of the (RE)BCO single-domain, which are correlated to the degree of seed dissolution, can be described (Figure 9): *(a)* a random growth mode: the seed dissolves completely, *(b)* an incompletely growth mode: the seed dissolves partially and resolidified leading to secondary nucleation events and *(c)* a well controlled growth: the seed does not dissolve leading to the formation of a single-domain.

In the second case, when the seed dissolves partially, the dissolved parts tent to re-solidify at low temperature which leads to a difficult control growth mode of the (RE)BCO single-domain at the seed. Subsidiary misoriented RE-123 grains are growing in the vicinity of the seed as can be seen on figure 10(a). In this case the diffusion of rare earth atoms originating from the seed is limited as compared to the complete dissolution case. The region where the liquid contains both rare earth elements is spatially limited. In this case the dissolved part tends to re-solidify into (RE, RE )$Ba_2Cu_3O_{6+x}$ and (RE, RE )$_2BaCuO_5$ phases preferentially at the surface of the undissolved part of the initial seed during cooling, which leads to a more or less controlled growth mode of the RE-123 grains at the seed [80].

In the third case, when the seed does not dissolve during the thermal treatment, a well-controlled growth mode of the (RE)BCO single domain can be observed (figure 10(b)).

The seed is partially soluble in the melt formed at high temperature by the peritectic decomposition of the RE-123 phase from the bulk (see figure 1). The solubility of the seed increases with increasing processing temperature and processing time [61]. Jee *et al.* [61] have studied the dissolution kinetics of the Sm-123 seed crystal. They found two independent factors affecting the dissolution kinetics of Sm-123 seed: *(i)* holding time at high processing temperature and *(ii)* chemical compositions of seed and compact.

In the hot seeding process [82,83], the degree of Sm-123 seed dissolution is limited [61]. Since the seeding is performed at the end of the holding period, the holding time at high temperature is relatively shorter than for a cold seeding process.

By controlling the chemical compositions of seed and compact, a minimization of the seed dissolution can be obtained. Melt-textured seeds are less dissolved than single-crystal seeds obtained by the flux method [84-88]. With melt-textured seeds, due to the presence of RE-211 particles in the seed and thus at the interface seed-compact, the contact area between the seed and the liquid phase is limited, which significantly decreases the dissolution rate of the seed, and hence produces well-controlled Y-123 crystals [89].

According to Kim *et al.* [62], the thickness of the seed is another factor, responsible for the partial or total seed dissolution. The authors also showed that a critical seed thickness is necessary for growing YBCO single-domain without formation of subsidiary grains.

Based on our recent investigations [90] in order to avoid the Nd-123 single-crystal seed dissolution during the TSMTG process of DyBCO single-domains, we conclude that it is useful to saturate the liquid phase formed during the peritectic decomposition reaction with Nd. To achieve this Nd saturation in the liquid phase, Nd-422 is added *ab initio* in the compact. Here the solubility of the seed is suppressed due to an excess neodymium content in the liquid phase. The as-obtained single-domain does not present subsidiary misoriented grains even if the thickness of the seed is lower than the critical thickness proposed by Kim *et al.* [62], or the processing time is longer than usual.

One of the main problems in the fabrication of high-quality (RE)BCO single-domains is the nucleation of secondary grains in the compact, which severely limits the growth of the main domain (from the seed) and results in a reduction of the final size of the single-domain. A small temperature window exists between the temperature where the heterogeneous nucleation is promoted from the seed and the temperature where an homogeneous nucleation appears [91,89]. Figure 11 shows the surface nucleation on the top of a DyBCO compact. It can be seen that the growth of the seeded single-domain is avoided by the formation of additional grains ahead of the growth front. The sample becomes multidomains.

It is thus important to carefully control the processing temperature in order to keep the system in this temperature window [12,57,65-67,92,93]. The final microstructure will be strongly affected by the cooling rate of the sample toward the lower temperature for crystallization [65,67,93].

Nevertheless, small thermal inhomogeneities (uncontrollable thermal gradients) often exist in the furnaces and may cause secondary nucleations that limit the growth of the single-domain. Controlled imposed thermal gradient [64] or the use of a cold finger near the seed [94,66] can help to prevent secondary nucleations. But the scaling up of the production is really difficult when a controlled thermal gradient must be applied. Batch production is easier when synthesis is realized in isothermal conditions. Shi *et al.* [65] reported that the secondary nucleation can be prevented by coating the sample surface with low melting compounds like CuO or silver. CuO coating leads to a local reduction in the homogeneous nucleation temperature of the system. They also found that coating the surface with a thin layer of RE-211 powder helps to prevent secondary nucleations and significantly increases the growth rate of the central domain. Meignan *et al.* [94] proposed to coat the bulk with an $Yb_2O_3$ slurry. $Yb_2O_3$ coating reacts with the melt to form $(Yb_{2-x}Y_x)BaCuO_5$ and an Yb-containing liquid phase, which reduces locally the peritectic temperature below 1010¡C (for Y-123) and therefore limits secondary nucleations.

Heterogeneous nucleations also appear at the compact/substrate interface and severely limit the thickness of the single-domain. Moreover, the Ba- and Cu-rich liquid phases formed at high temperature by decomposition of the RE-123-phase is highly corrosive. This liquid reacts with most of the commonly used crucibles and substrate materials. These reactions have to be minimized as they deteriorate the superconducting properties of the samples due to impurity diffusion into the RE-123 crystal lattice. Moreover, separation of the single-domain from the substrate is difficult at the end of the melt processing [94].

MgO single-crystal substrates were often used as they are relatively inert to the melt because (001) MgO plane is not wetted by the liquid phases [95,57]. The MgO substrate used for melt processing is easily removable from the single-domain after the melt process and could be used repeatedly. The disadvantage of the MgO substrate stems from the fact that RE-123 grains nucleate on the compact/MgO substrate interface, which makes it difficult to fabricate a single-domain sample [57]. Another disadvantage is that MgO single-crystals are rather expensive and thus the availability of large size (5-10 cm) (RE)BCO pellets is limited [94].

The RE-123 grains rarely nucleate on $ZrO_2$ and $Al_2O_3$ substrates but these react significantly with Cu and Ba-rich liquid phases to form $BaZrO_3$ and $BaAl_2O_4$ respectively [32,96,97]. These reactions severely damage the superconducting properties.

It has been found that the most effective way to suppress surface nucleation and particularly compact/substrate nucleation is to coat the sample surface with again low melting compounds like Yb-123 or $Yb_2O_3$ [57,94,67]. $Yb_2O_3$ forms with BaO and CuO at high temperature a liquid phase at the compact/substrate interface. The Yb-123 phase has the lowest peritectic temperature ($T_p$=980°C[94]) of the RE-123 family compounds. Therefore the peritectic temperature of the liquid phase at the compact/substrate interface is between 980°C and the peritectic temperature of the RE-123 phase in the bulk. Thus, the liquid containing ytterbium will be last solidified during peritectic cooling, suppressing the RE-123 nucleation at the compact/substrate interface [60]. The suppression of secondary nucleation by coating Y-123 compact with $Yb_2O_3$ was thoroughly investigated by Kim *et al.* [57]. They observed that in the $Yb_2O_3$-coated region, many Yb-211 particles were formed by a reaction between $Yb_2O_3$ and the liquid phase from the compact. They also found CuO phase at the interface between the compact and $Yb_2O_3$ coating, due to the fact that the mass balance in this region could not satisfy the RE-123 phase formation conditions. The reactions that take place in this region, induce local off-stoichiometric composition and therefore affect the growth process of the single-domain. However, the ytterbium diffusion is not very fast and its effect on the microstructure of the interior part of YBCO compacts is not significant. When $Yb_2O_3$ coats the compact, the sample can be processed on polycrystalline reactive substrate materials like $Al_2O_3$ rather than on more expensive inert single crystal substrates. As compared to Yb-123 coating, the oxide coating provides some advantages as it can react with the liquid phase and form RE-123, thereby reducing the loss of liquid phase [94].

In conclusion seed dissolution, surface and compact/substrate nucleation and contamination which limit the fabrication of large (RE)BCO single domains, can be overcome by optimising processing parameters such as the maximum processing temperature and holding time at this temperature, degree of undercooling, compositions of the seed and the compact, surface coating of compact,

## 3. Pores and cracks

Many defects such as pores and macrocracks are created during the heating and cooling step of the single-domain synthesis process. Some defects are also introduced during oxygen annealing.

On figure 12(a) spherical pores are observed within the DyBCO single-domain. The formation of pores is due to gas (oxygen) evolution during the incongruent melting of the RE-123 phase [19,25,26]. At high temperature, the RE-123 phase decomposes peritectically into RE-211 solid particles and a Ba- and Cu-rich liquid phase. During this incongruent melting some oxygen gas is released, forming spherical pores in the liquid [19] due to the reaction:

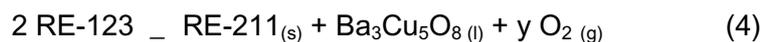

$$2 \text{ RE-123} \rightarrow \text{RE-211}_{(s)} + Ba_3Cu_5O_{8\ (l)} + y\ O_{2\ (g)} \quad (4)$$

The size and amount of pores depend on the heating rate to attain the peritectic temperature[58,19,56]. Larger pores are observed for higher heating rate. The influence of the heating rate on pores size and density in YBCO single-domain has been studied by Lo *et al.* [58]. They found that pores of about 1 to 2˚mm in diameter are created, when the heating rate is as large as 30¡C/h, due to the insufficient time for the gas to escape from the compact. For this reason, the density of pores is higher in the center of the bulk than on the sides. A reduction in the heating rate to 10¡C/h decreases the average pore size to 50 _m. However, further reduction in the heating rate below 10¡C/h results in a loss of liquid phase and leads to significant coarsening of RE-211 particles, which later may give a deleterious effect on the superconducting properties [19]. It was found that the presence of pores does not alter the growth direction of the RE-123 single domains, although they interrupt the *a-b* planes at the surface of the pore [56].

Even if pores do not alter significatively the superconducting properties of single-domains, macro-cracks in the material can strongly affect these properties. A multi-grain feature in trapped field profiles is often observed when macro-cracks cross the sample even if the sample looks like a single-domain [28]. Macro-cracks along the a/b planes can be observed on figure 12(a) and (b).

Cracking is formed because of thermal stresses during the thermal profile, in particular for large grain samples. The tetragonal to orthorhombic phase transformation also causes cracking mainly along the (001) cleavage planes [28,29]. Moreover, the macro-cracks have a high probability to propagate when the sample is submitted to a strong magnetic field [98,5] or to many thermal cycles (cooling below $T_c$ and heating up to room temperature) [99]. We have studied crack propagation in single-domains and have shown that the microstructure, mainly the RE-211 particle distribution, is a key parameter allowing us to improve the fracture toughness of samples [48]. It can be observed on figure 12(b) that macro-cracks propagate in the RE-211-free area (left side of the picture) and are stopped at the beginning of the area containing RE-211-particles (right-side of the picture).

Macro-cracks formation is strongly related to microstructure features. For example, a non-uniform distribution of RE-211 or large pore density can cause internal thermal stresses, due to differences in thermal expansion coefficients of the various phases, during the heat treatment [28].

Diko *et al.* [29] have studied the effect of oxygen annealing process on macro-cracks formation, like *a/b* macro-cracks formed during the oxygenation process. These cracks provide a path for oxygen to diffuse through the material. This causes a faster transformation to the orthorhombic phase along the cracks and consequently enhances crack propagation [29]. The authors observed that the cracks spacing increases with increasing the annealing temperature. The reason for cracking is suspected to be related to the evolution of the *c*-lattice parameter of the RE-123 phase with oxygen content [29].

## 4. Influence of RE-211 particles on microstructure of (RE)BCO single-domain

It has been mentioned that in melt-textured growth processes, the RE-123 domain growth is promoted by peritectic recombination of the RE-211 particles and the Ba and Cu-rich liquid phases. The rare earth ions needed for the growth of RE-123 phase are provided by the dissolution of RE-211 particles in the liquid. This dissolution is relatively slow with respect to the peritectic recombination process and leads to the presence of trapped RE-211 particles in RE-123 single-domain matrix as well as residual solidified liquid phases. These residual solidified liquid phases dramatically reduce superconducting properties. In order to consume completely the liquid phase during the growth process, extra RE-211 particles are added in the preform material. Generally around 20-30˚mol% of extra RE-211 particles are added in the system [11,24,89]. On the contrary to the residual solidified liquid phases, RE-211-particles in the final single-domain have a positive effect on superconducting properties, as the RE-123/RE-211 interfaces are suspected to be good pinning centres [20-24]. It was found that critical current density ($J_c$) increases with the $V_{211}/d_{211}$ ratio, where $V_{211}$ is RE-211 volume fraction and $d_{211}$ is the average size of RE-211 particles [100,20].

Moreover, addition of RE-211 particles allows to reduce the loss of liquid phase [89] and plays also a significant role on the mechanical properties of single-domains as it will be shown later [48,19,101].

Therefore the control of the homogeneity and size distribution of RE-211 particles in the RE-123 single-domain matrix is very important to improve the quality of single-domains [48].

### *4.1 pushing/trapping theory*

When the single-domain is growing, phenomena similar to a pushing/trapping behaviour are observed and it results some inhomogeneities in the RE-211 particle distribution. The pushing/trapping has been widely considered by many investigators in various materials. Endo *et al.* [79,102] applied the pushing/trapping theory to the YBCO system by considering inactive particles (Y-211) at an advancing solid/liquid interface (Growth front of Y-123) during solidification. This interaction causes forces on RE-211 particles resulting in pushing them along the solidification front or trapping them in the RE-123 solid phase. As shown on figure 13, two dominant forces acting on a RE-211 particle were considered:

(i) A drag force ($F_d$), due to the viscous flow around the particle, which always leads to trapping the RE-211 particle.

(ii) A force ($F_i$) due to the interfacial energy ($\Delta\sigma_0$), which occurs as the interface approaches the particles closely enough.

The interface energy ($\Delta\sigma_0$) in the RE-123/RE-211 system is defined as:

$$\Delta\sigma_0 = \sigma_{sp} - \sigma_{lp} - \sigma_{sl} \qquad (5)$$

where $\sigma_{sp}$, $\sigma_{lp}$, and $\sigma_{sl}$ are the solid/particle, liquid/particle, and solid/liquid surface energies, respectively.

It was shown from this theory [79,102] that the probability of trapping a RE-211-particle in the growth front depends on the size of the particle, the rate of the growth front and the interfacial energy ($\sigma_0$). Therefore, at a given growth rate, the particles with size smaller than a certain critical radius $r^*$ are not trapped in the growing RE-123 solid phase. The relation between growth rate $R$, critical radius $r^*$ and interface energy $\sigma_0$ is given by

$$R \propto \frac{\Delta \sigma_0}{\eta \, r^*} \qquad (6)$$

where $\eta$ is the viscosity of the melt.

The growth rate, $R$, is proportional to the undercooling, $\Delta T$ [63]. When $\Delta T$ increases, the critical radius for trapping decreases as the growth rate increases, and small RE-211 particles can be trapped. When the sample is cooled down slowly, the undercooling and the growth rate increase with the distance from the seed. At the beginning of the solidification process, $\Delta T$ and $R$ are relatively low; therefore small RE-211 particles are pushed by the growth front. Only later at higher undercooling, trapping into the RE-123 solid phase occurs (see figure 14). Consequently there is an increase of RE-211-volume fraction, $V_{211}$ with the distance from the seed and a decrease of the size of RE-211 particles, $d_{211}$ [100]. The critical RE-211-particle radius depends not only on the growth rate, which is lower in the $c$-growth direction than in others, but also on the solid-liquid interface energy, $\sigma_0$. Endo et al. [79] assume that $\sigma_0$ is larger for (001) plane (*i.e.* along de $c$-growth direction) than for (100) plane (*i.e.* along $a$-growth direction) in Y-123 crystal. Therefore for the same undercooling, the critical radius of RE-211 particle for $a$-direction growth is smaller than for $c$-direction growth, $r^*_a < r^*_c$ due to growth rate anisotropy and interfacial energy differences for $a$- and $c$-growth directions. Thus, more RE-211 particles are trapped for $a$- growth direction.

Even if this theory is in relatively good agreement with the experimental results, one should be careful when applying the pushing/trapping theory to the RE-123/RE-211 system because this theory was initially proposed for inactive inclusions in the solid and non-facetted materials. In fact RE-211 particles are active inclusions because they supply the system with rare-earth ions from RE-211 dissolution in the melt. Moreover the RE-123 anisotropically grows with facets.

For example, RE-211-particles pushed out by the growing RE-123-crystal are expected to become smaller by self-decomposition in the process for supplying rare-earth ions and/or become larger due to Oswald ripening [79].

## *4.2. RE-211-free regions*

Figure 15 shows some spherical RE-211-free regions. These inhomogeneities in the RE-211 distribution have been associated with the formation of spherical pores due to oxygen gas evolution during melt-textured process [23,101,19,103]. When the oxygen diffuses out of the pores, these are filled by a liquid phase and produce spherical liquid pockets containing a few RE-211 particles. During the slow cooling in the melt-texturing process, the liquid pockets are transformed into the RE-123 phase, which do not contain RE-211 particles [103]. Due to the relatively slow mobility of the RE-211 particles as compared with the liquid motion in the pores it was observed that a non-uniform distribution of RE-211 particles occurs around the liquid pockets. The reaction to form the RE-123 phase is sometimes difficult in these regions due to the low RE-211 concentration around the liquid pockets. Therefore an unreacted liquid phase was often observed in the centre of these RE-211-free regions [103].

## *4.3. RE-211 Coarsening*

As we already noted before, the presence of RE-211-particles in the final single-domain has a positive effect on the critical current density. This is due to the fact that the RE-123/RE-211 interface is suspected to be a good pinning centre [20-24, 104]. Therefore for a given volume fraction of RE-211 particles, the RE-123/RE-211 interface is maximized when RE-211 particles are small. However, in the classical Top-Seeded Melt-Textured growth process, the RE-211 particles have the tendency to be relatively big. Indeed, most of the RE-211 particles are formed at high temperature by peritectic decomposition of the RE-123-phase. Under these conditions, nucleation and growth of these particles is extremely difficult to control. Moreover, the RE-211 particles are subjected to the Oswald ripening phenomenon at high temperature in the melt [105]. In the Oswald ripening process, the smaller RE-211 particles are dissolved in the melt, and dissolved ions migrate to grow the biggest particles.

In order to improve $J_c$, it is very important to overcome the RE-211 coarsening problem. Several techniques for reducing the RE-211 size, such as lowering the processing temperature as well as shortening the delayed time at high temperatures [106], using small-sized precursor powders [107,108] and adding coarsening inhibitors [105,109-111], have been developed. The common coarsening inhibitors additives are Pt [11,47,15], $PtO_2$ [112]; $CeO_2$ [34,38,105,111,113-115], $SnO_2$ [30,89]; $ZrO_2$ [32], $BaZrO_3$ [32,116]. Most of the time, these

additives also change the morphology of the RE-211 particles. It has been found that the Pt- and Ce-additions are the most effective ones [11,115,105].

*a) Pt-additives*

There are many studies which reported an enhanced critical current density $J_c$, in Pt-doped (Pt or $PtO_2$) RE-123 single-domains [47,112,34,15]. It was proposed that such effects might be associated with the fine, uniform dispersion of RE-211 particles resulting from the addition of Pt.

Two different mechanisms have been proposed in order to explain the role of the Pt-compounds on the RE-211-particles refinement. The first one is an heterogeneous nucleation on Pt containing particles [117]. The other one is related to the modification of the RE-211/liquid interfacial energy by the Pt addition. The latter role seems to be the preponderant one and strongly limits the coarsening of RE-211 particles in the liquid [118].

*b) $CeO_2$ additives*

The $CeO_2$ addition has been considered as an effective and cheap additive to yield fine RE-211 particles in melt-textured RE-123 domains. Numerous studies have been performed on the understanding of the refinement mechanism of $CeO_2$ addition. Three possible roles of the impurity element on the refinement of RE-211 have been proposed [119]: *(i)* a nucleation assistant; *(ii)* a grain growth inhibitor and *(iii)* a modifier of interfacial energy of RE-123/RE-211.

Pinol *et al.* [109] have suggested that $CeO_2$ reacts with Y-211-precursor in the YBCO system to produce a $BaCeO_3$ phase and nanometric $Y_2O_3$ particles. (eq 6)

$$RE\text{-}211 + CeO_2 \rightarrow BaCeO_3 + CuO + RE_2O_3 \quad (7)$$

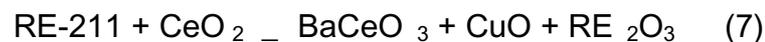

They have proposed a heterogeneous nucleation mechanism in which Y-211 particles nucleate on nanometric $Y_2O_3$ particles due to $CeO_2$ addition. Chen *et al.* [38,111] shown that about one-fourth Y-211-particles contain at least one nanometric $Y_2O_3$ core. In addition, they observed 1°m square $BaCeO_3$ particles dispersed in the bulk. These observations are in a good agreement with the explanations given by Pinol *et al.* [109]. New Y-211-particles are formed around the $Y_2O_3$ by consuming the over-saturated yttrium in the liquid phase. Therefore the possible coarsening of the existing RE-211-particles is limited [111].

On the other hand, Kim *et al.* [19] have used a liquid infiltration process to point out that $CeO_2$ reacts with the Y-123 phase rather than the Y-211 phase and plays the role of growth inhibitor. They have argued that the Y-211 particles containing $Y_2O_3$ cores in textured YBCO might be a special case, not a general one.

$$2\ RE\text{-}123 + 3\ CeO_2 \_ 3\ BaCeO_3 + 5\ CuO + RE\text{-}211 \qquad (8)$$

### *4.4. Microstructural defects due to the presence of RE-211 particle in RE-123 single-domain*

It was mentioned before that microstructural defects such as dislocations act as magnetic flux pinning sites in (RE)BCO single-domains and enhance $J_c$. In single-domains other microstructural defects such as microcracks or stacking faults are also created by the presence of RE-211 particles. The microstructural defects also play a significant role in the magnetic field pinning process.

*a) Micro-cracks*

Microcraks can be observed in optical and scanning electron microscopy as characteristic lines parallels to the *a-b* planes forming the platelet structure of the RE-123 grain (see figure 16). Microcracking thus appears as an important phenomenon in melt-processed (RE)BCO superconductors. Diko [120] reported the origin of microcracking in (RE)BCO materials by taking into account the differences in thermal expansion coefficients and in elastic constants for RE-123 and RE-211 phases. From such considerations, a description of the thermal stresses arising in the RE-123 phase during cooling has been proposed. A critical RE-211 particle size for inducing *a-b* microcracking has been estimated using a simplified model [120,121].

*b) Stacking faults*

Other types of defects observed near the RE-123/RE-211 interfaces, are stacking faults. Stacking faults can be explained as an additional Cu-O layer located between two Ba-O layers or by the intergrowth of a RE-124 phase in the RE-123 crystal matrix [122]. It was found that stacking faults are formed during oxygen annealing. Indeed, during the oxygen annealing process the RE-123-phase is unstable and could decompose into other stable phases, one of these decomposition reactions generates CuO. In the case of melt-processed samples, this RE-123-decomposition mainly occurs at the RE-211/RE-123 interface. This can be associated with stresses induced by *(i)* the tetragonal-orthorhombic transformation and *(ii)* the difference of thermal expansion coefficients between RE-211 and RE-123 phases [19].

Stacking faults thus appear around the trapped RE-211 particles within the RE-123 matrix [123].

It was also found that $Y_2O_3$ formed due to the low stability of the RE-123-phase in the annealing process leads to the formation of other types of stacking faults - $Y_2O_3$ stacking faults [62]. The density of stacking faults increases with increasing oxygenation annealing time. It was reported that the stacking faults near the RE-211/RE-123 interface have a width of a few tens of nanometres and a length of a few hundred nanometers [110]. Therefore, these stacking faults can act as pinning centres and enhance critical current density $J_c$.

## 5. Influence of RE-211 particles on the mechanical properties

From considerations based on the thermal and elastic mismatch effects between Y-211 and Y-123 phases, and from microstructural data, Goyal and co-workers [124] have argued that the Y-211 particles can be considered like a reinforcement material contributing to the enhancement of the fracture resistance behavior of Y-123 matrix through energy dissipation due to interfacial delamination and crack bridging. The former mechanism of energy dissipation may result from the strong chemical bonding existing between both phases when a high modulus reinforcement particle is added to a lower modulus matrix[125]. Measurements made by microindentation technique indicate that Y-211 particles are characterized by a higher average Young modulus (213 GPa) than the Y-123 matrix (180GPa) [124].

Thermal shock resistance has been also improved by the presence of RE-211 particles up to a certain degree of RE-211 concentration, the RE-211 particles acting as solid links between RE-123 cracked layers. Because of the large difference between thermal expansion coefficients of RE-123 and RE-211 phases, residual thermal stresses are created around RE-211 particles on cooling leading to a strong microcracking of the RE-123 matrix along the (001) planes [126].

In our group, we have studied in details [48] the propagation of cracks in DyBCO single-domains by SEM analysis. Two significant roles of the Dy-211-particles were put into evidence.

First, it can be seen (figure 17(*a*)), that the crack deviates from its initial path, when it reaches a Dy-211-particle (in white in the micrographs), and tends to propagate around the Dy-211 particle, due to the residual mismatch stress field around them. This can be related to the difference of Young modulus between RE-211-particles and RE-123-matrix: the incorporation of high modulus particles in a low-modulus matrix results in a reduction of the tensile stress at the tip of a crack running through the matrix [124]. It is clear from these observations that the Dy-211-particles play a first significant role in the Dy-123 toughening mechanism by deviating the cracks.

Another possible mechanism of toughening was put into evidence in these materials, and is illustrated in figure 17(*b)* and *(c)*. In areas characterized by a high concentration of RE-211 particles, a wake bridging phenomenon is observed. Thus RE-211 particles play another retardant role in the crack propagation mechanism by crack bridging, resulting in an increase of the fracture toughness of the material.

Therefore, crack propagation in RE-Ba-Cu-O strongly depends on the RE-211 particles distribution in the matrix, which is not homogeneous [48]. It is thus extremely important to control the distribution of these particles in the RE-123-matrix in order to improve the mechanical properties of single-domains.

The addition of silver, which is permeable to oxygen, into the matrix may improve drastically the mechanical properties, reaching a toughness value, $K_c$, equal to 1.88 MPa m$^{-1/2}$, without affecting the superconducting properties [49,127-130].

It is also possible to impregnate the bulk with epoxy resin [131]. This allows avoiding the crack extension favoured when the superconductor is placed repeatedly under a magnetic field. The epoxy resin can penetrate into the material micro-cracks from the surface and fill pores that are connected to the cracks. Moreover, this resin also avoids the exposure to moisture and thus to corrosion.

## IV. Alternative process for controlling the RE-211 distribution in RE-123 single-domain: The Infiltration and Growth Process

As we discussed in the previous paragraphs, the control of the RE-211-particles distribution is an important problem in the manufacture of YBCO-type single-domains [16,23,53,54,79,101,104,115,116,121,132]. The usual TSMTG process leads to samples with large size and heterogeneously dispersed RE-211-particles and of various sizes. Moreover, the morphology of these particles cannot be easily controlled due to the fact that most of the RE-211 particles directly result from the peritectic decomposition of the RE-123-phase at high temperature, and the others result from the pro-peritectic RE-211-particles inserted at the beginning in the green body.

The Infiltration and Growth Process has been proposed by Reddy *et al.* [45,133] to be an alternative technique for single-domain synthesis allowing for controlling the RE-211 particles distribution. This process is based on the infiltration of a barium and copper rich liquid phase into a RE-211 preform at a temperature near (below or above) the peritectic temperature of the RE-123 phase. We have studied [137] the main parameters (set ups, maximum processing temperature with respect to the peritectic temperature, nature of reactant, porosity of the RE-211 preform) of the Infiltration and Growth Process. Moreover, we have proposed different processes of *chimie douce* in order to produce Dy-211 particles with controlled shape and size, that can be used as precursors for the Infiltration and Growth Process

Figure 18 shows a comparative scheme of TSMTG and Infiltration & Growth processes. This technique allows one to obtain high density materials with small size RE-211-particles well dispersed in the bulk [134]. Moreover a near-net shape material can be obtained. Indeed, the process leads to a reduced shrinkage which limits cracks and distortions of the matrix [45]. It is thus possible to manufacture samples with complex shapes. For example, Reddy *et al.* [135,136] have used the technique to manufacture $YBa_2Cu_3O_{7-d}$ superconductor fabrics starting from $Y_2O_3$ clothes. This process was also used by the same authors [44] to produce YBCO superconducting foams. In this case, polyurethane foams with controlled porosity were coated with a Y-211 slurry before being burned out. The resulting Y-211 foams were then infiltrated with a Ba- and Cu-rich liquid phase and seeded with a Nd-123 single-crystal in order to produce a new kind of single-domain with controlled high porosity content. A home-made Dy-123 superconducting foam can be seen on figure 19.

## V. CONCLUSION

Superconducting and mechanical properties of YBCO-type single-domains are strongly dependant on the microstructure. The synthesis process influences the microstructure of the as-obtained samples. To be able to control the growth, it is necessary to prevent the seed dissolution. Different kind of seed, with different advantages, can be used. It is also important to avoid parasite nucleation on the surface of the bulk and at the compact/substrate interface. These problems can be solved by carefully controlling the thermal cycle used for the growth and by using low melting point buffer layer (like $Yb_2O_3$) between compact and substrate.

During the melt-texturing process, pores and cracks are created in the sample. Cracks are also formed during the oxygen annealing process. Pores do not alter the growth of the single-domain and the superconducting properties. On the contrary macro-cracks damage the superconducting properties of the samples.

In this review we have focused our attention on the presence of RE-211 particles trapped in the RE-123 matrix, since RE-211 particles play a significant role on the superconducting and mechanical properties of the single-domains. Phenomena like pushing-trapping and Oswald ripening were discussed. Different additives to limit the coarsening of the RE-211 particles, like Pt and $CeO_2$, were proposed.

Formation of microstructural inhomogeneities, like regions with low content of RE-211-particules, called RE-211-free areas, was treated

Micro-cracks and stacking faults are microstructural defects caused by the presence of RE-211-particles in the RE-123-matrix. These defects are found to play a beneficent role in the magnetic field pinning process.

Mechanical properties of single-domains, and particularly toughness, are improved by the presence of the RE-211-particles in the matrix. Cracks are deviated from their initial path when they reach a RE-211-particle. This can be explained by the difference of Young modulus between RE-211-particles and the RE-123-matrix. A wake bridging phenomena is also observed in areas with high concentration of RE-211-particles. Since RE-211-particles play a significant role in the toughening mechanism of (RE)BCO single-domains, it is extremely important to control the distribution of these particles in the RE-123-matrix, all the more as the microstructure of the single-domains obtained by the classical Top-Seeded Melt-Textured Growth process is deeply inhomogeneous.

Infiltration and Growth process was proposed as an interesting alternative to the classical TSMTG, since it is possible to control the spatial and size distribution of the RE-211 particles in the RE-123 single-domain matrix.

ACKNOWLEDGEMENTS

The preparation and the characterization of materials are parts of the doctorate thesis of J-P Mathieu who thanks FRIA (Fonds pour la Formation la Recherche dans l Industrie et dans l Agriculture), Brussels for financial support. Ms Koutzarova has been financially supported by the European Supermachines Research Training Network (HPRN-CT-2000-0036). Part of this work results from research activities in the framework of the VESUVE project of the Walloon Region (RW.01.14881).

## VII. FIGURE CAPTIONS

**Fig. 1:** Pseudo-binary phase diagram for the _$Dy_2O_3$-BaO-CuO system in the vicinity of the peritectic reaction (RE-123 _ RE-211 + Liquid).

**Fig. 2:** SEM micrography of DyBCO sample. Dy-211-particles (in bright color) are entrapped in the Dy-123 superconducting matrix (in dark gray).

**Fig. 3:** Optical micrograph of a DyBCO polygrain sample. Liquid segregation can be observed at the grain boundary.

**Fig. 4:** Optical micrograph of a DyBCO polygrain sample.

**Fig. 5:** SEM micrograph of a Nd-123 single-crystal.

**Fig. 6:** Photography of a YBaCuO single-domain sample.

**Fig. 7:** Top-Seeded Melt-Textured (TSMT) (RE)BCO single-domains generally grow in a parallelepipedic form with (100), (010) and (001) crystal habit planes.

**Fig. 8:** SEM micrography of the growth front.

**Fig. 9:** Three different growth modes of the RE-123 single-domain depending on the degree of dissolution of the seed.

**Fig. 10:** Samples photography with (a) partially dissolved seed that limit the control of the domain growth; (b) non-dissolved seed that allows a good single-domains to growth

**Fig. 11:** Surface nucleation on the top of a DyBCO compact

**Fig. 12:** Optical micrographies of DyBCO single-domains. (A) Pores can be observed on the bottom part of the picture. (B) Zoom of an area near the seed. It can be observed that macrocraks are present in the Dy-211-free area and stop at the beginning of Dy-211-containing area.

**Fig. 13:** Schematic drawing showing a particle in the front of the solid/liquid interface and the forces acting on the particles. $F_d$ and $F_i$ indicate the drag force due to viscous flow around the particle and the force due to interfacial energy, respectively

**Fig. 14:** Model of the RE-211 particle trapping phenomenon in the RE-123 phase for different undercooling. From ref. [79] and [102]

**Fig. 15:** SEM micrographies showing Dy-211 free area in a Dy-123 single-domain matrix. Dy-211 particles appear in bright gray on the picture and Dy-123 in dark gray.

**Fig 16:** Optical micogaphy of DyBCO single-domain. Microcraks can be observed as characteristic lines parallels to the a-b planes forming the platelet structure of the RE-123 grain

**Fig 17:** Crack propagation in the Dy-123-matrix: (a) Crack path modified by Dy-211-particles; (b) and (c) crack jump from one Dy-211-inclusion to another due to the high proximity between Dy-211-particles.

**Fig. 18:** Comparison between TSMTG process and IG process.

**Fig 19:** Photography of a DyBCO superconducting foam

## Figure 1

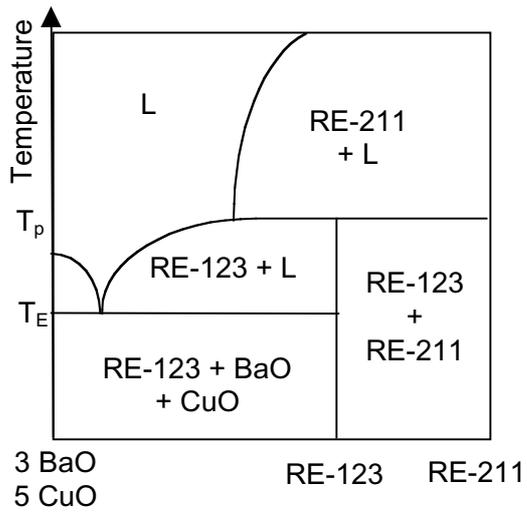

According to the peritectic decomposition:
(RE)-123 _ (RE)-211 + liquid

## Figure 2

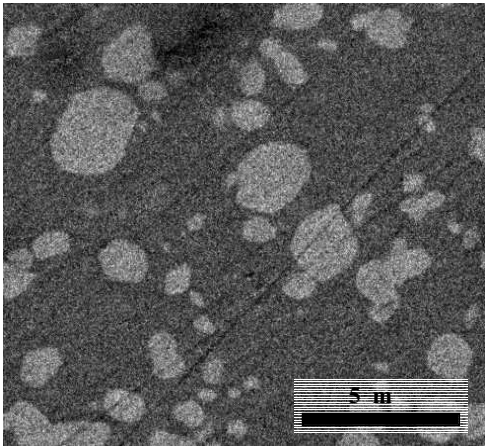

## Figure 3

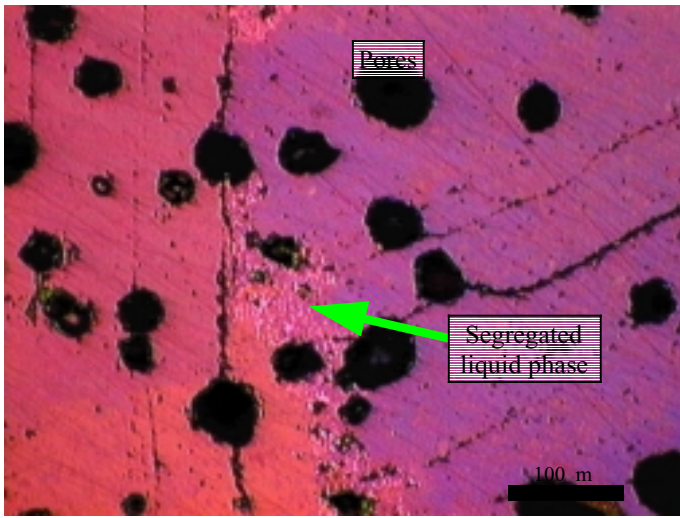

## Figure 4

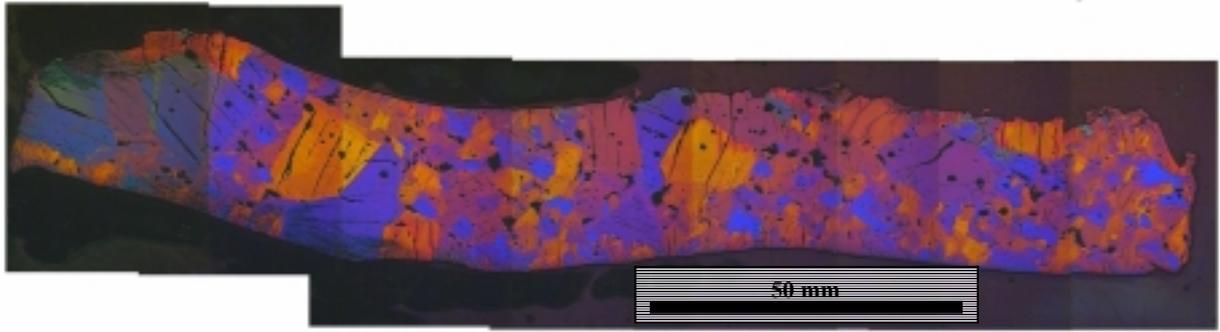

## Figure 5

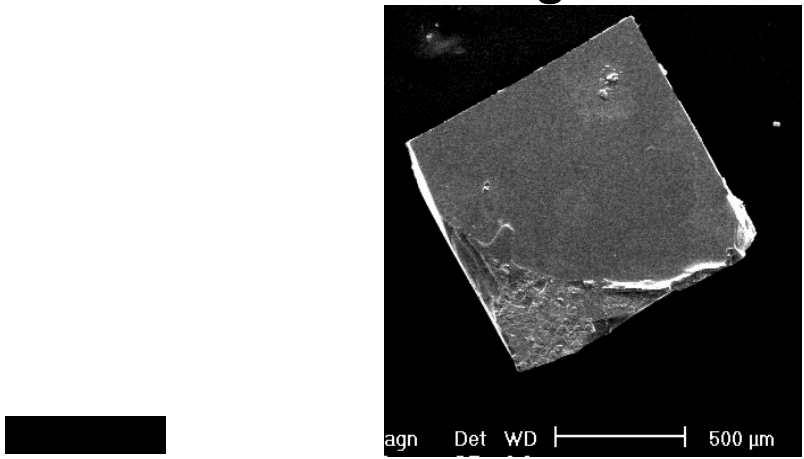

## Figure 6

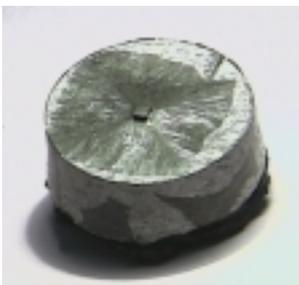

# Figure 7

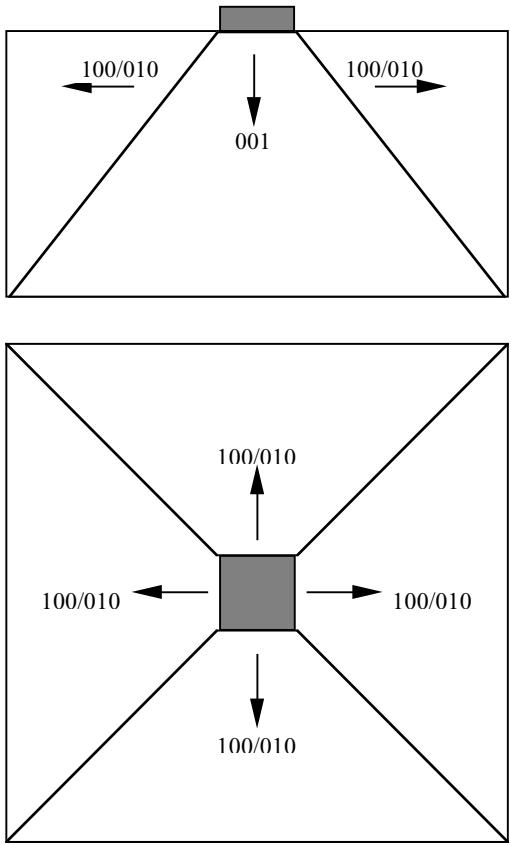

# Figure 8

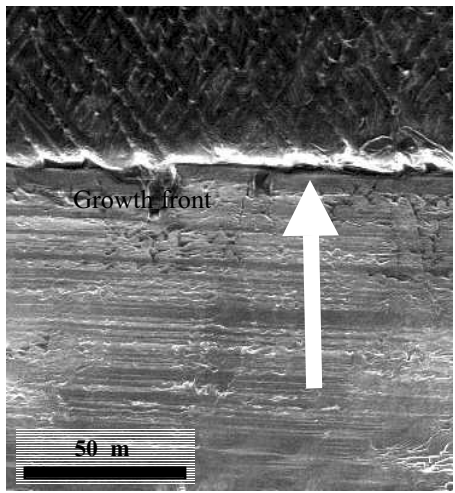

## Figure 9

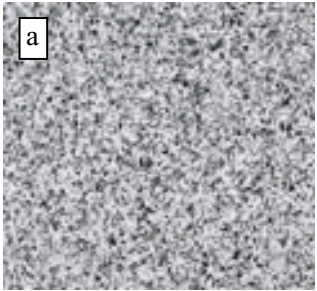 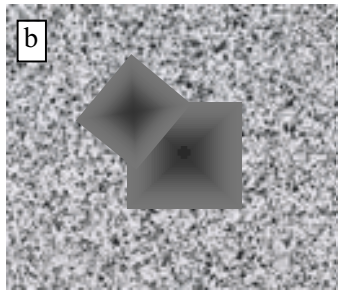 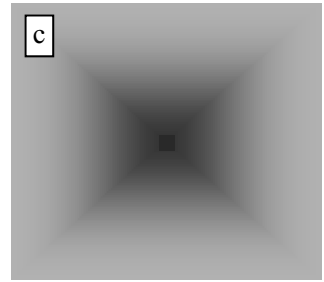

a. The seed dissolves completely

b. The seed dissolves partially and resolidified

c. The seed does not dissolve

## Figure 10

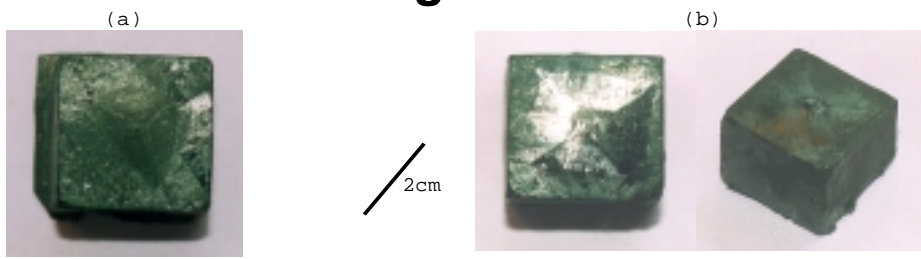

(a)   (b)

2cm

## Figure 11

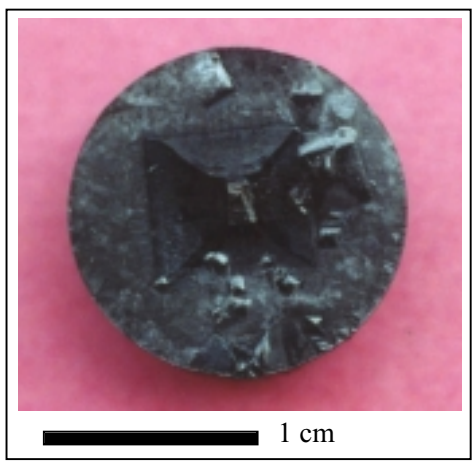

1 cm

## Figure 12

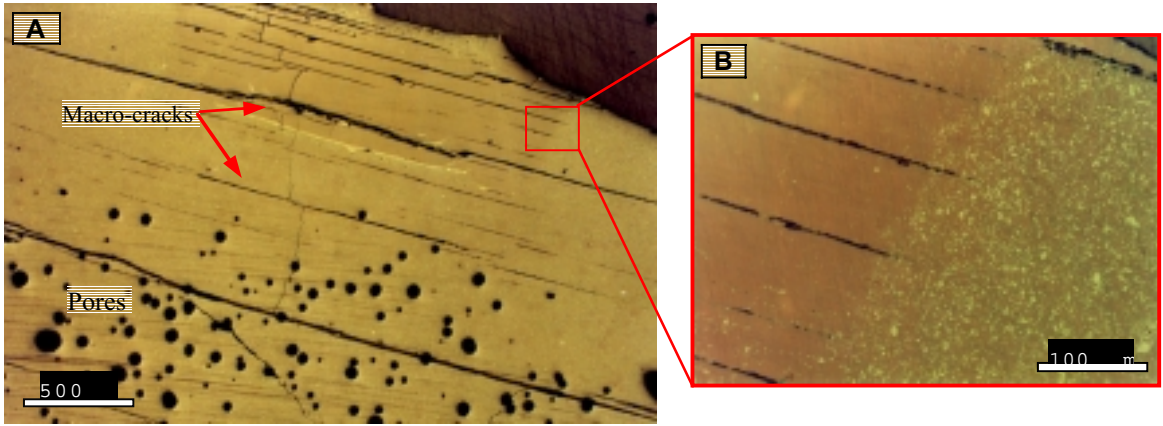

## Figure 13

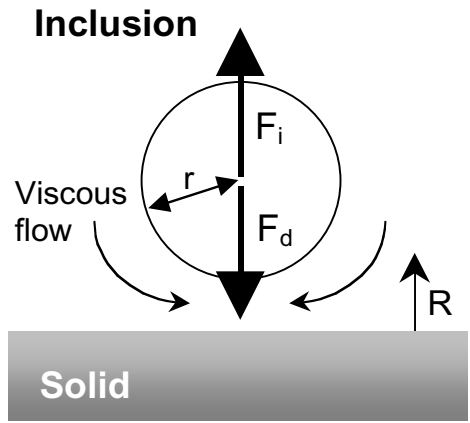

## Figure 14

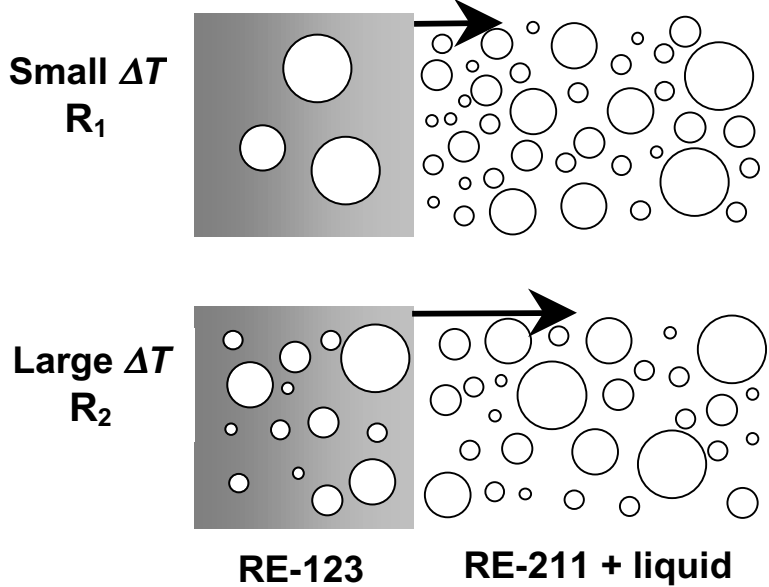

Figure 15

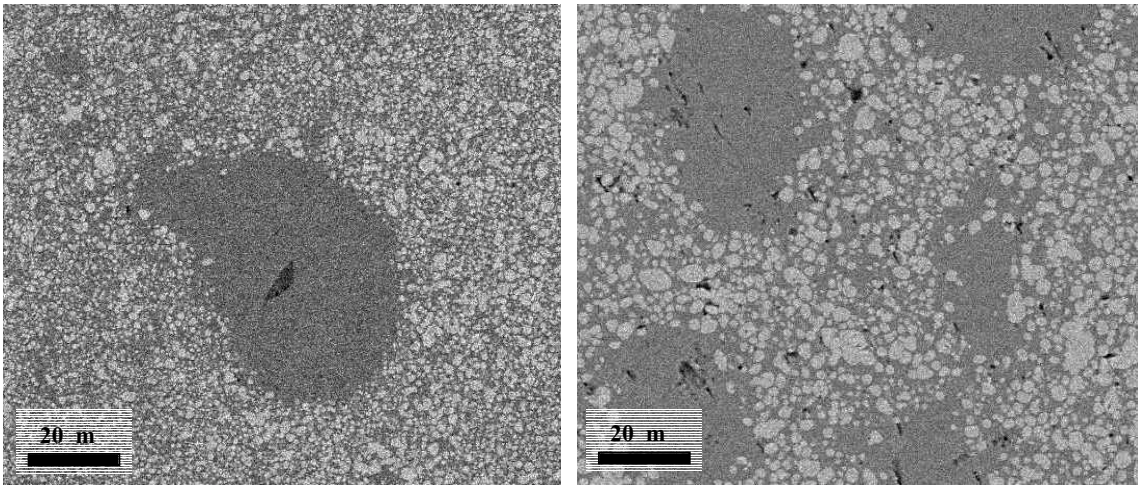

Figure 16

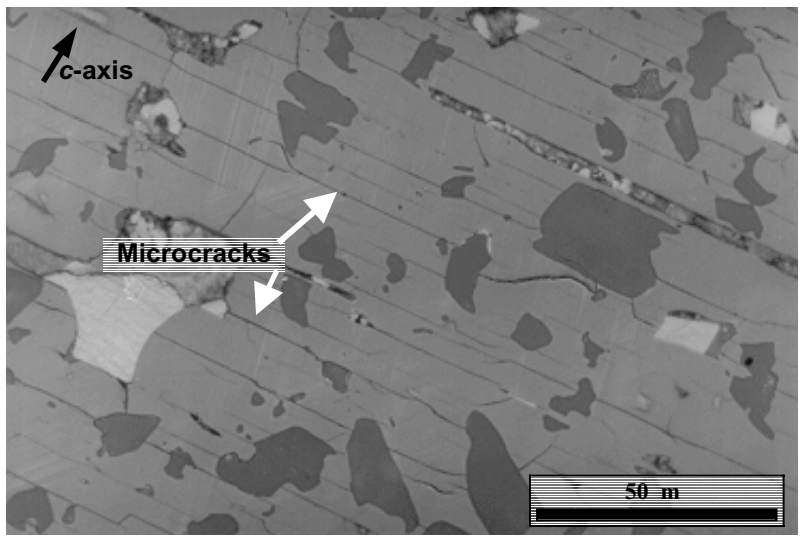

*Figure 17*

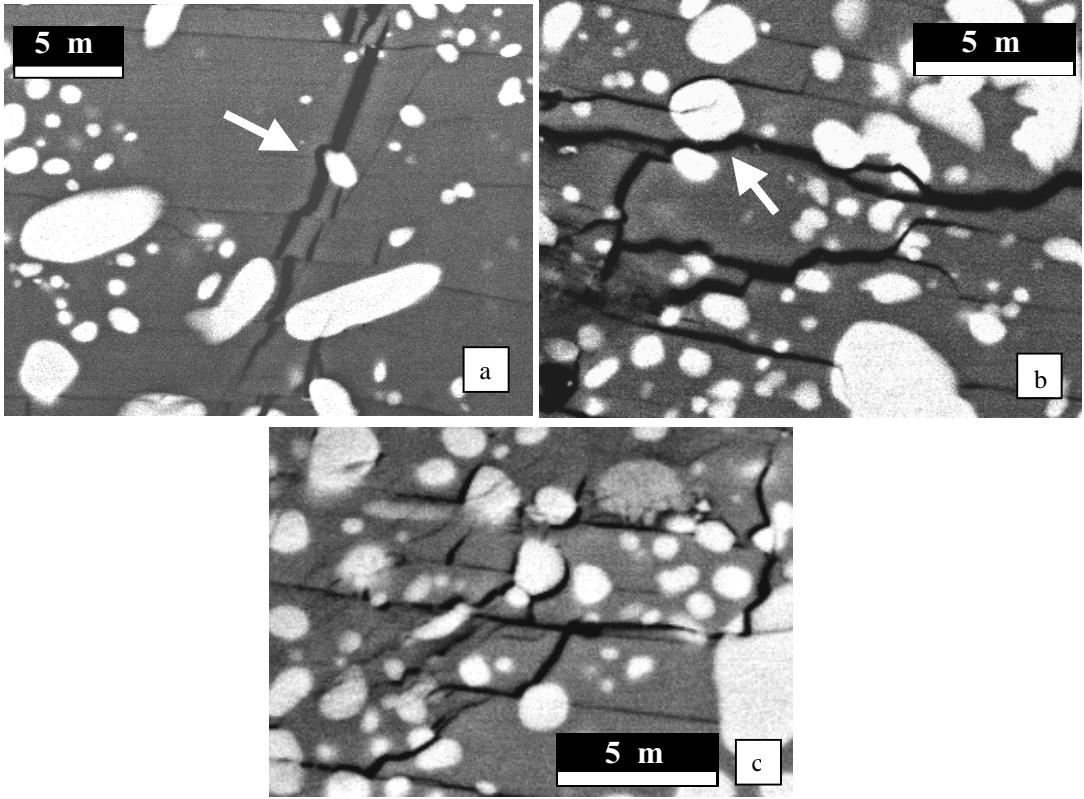

*Figure 18*

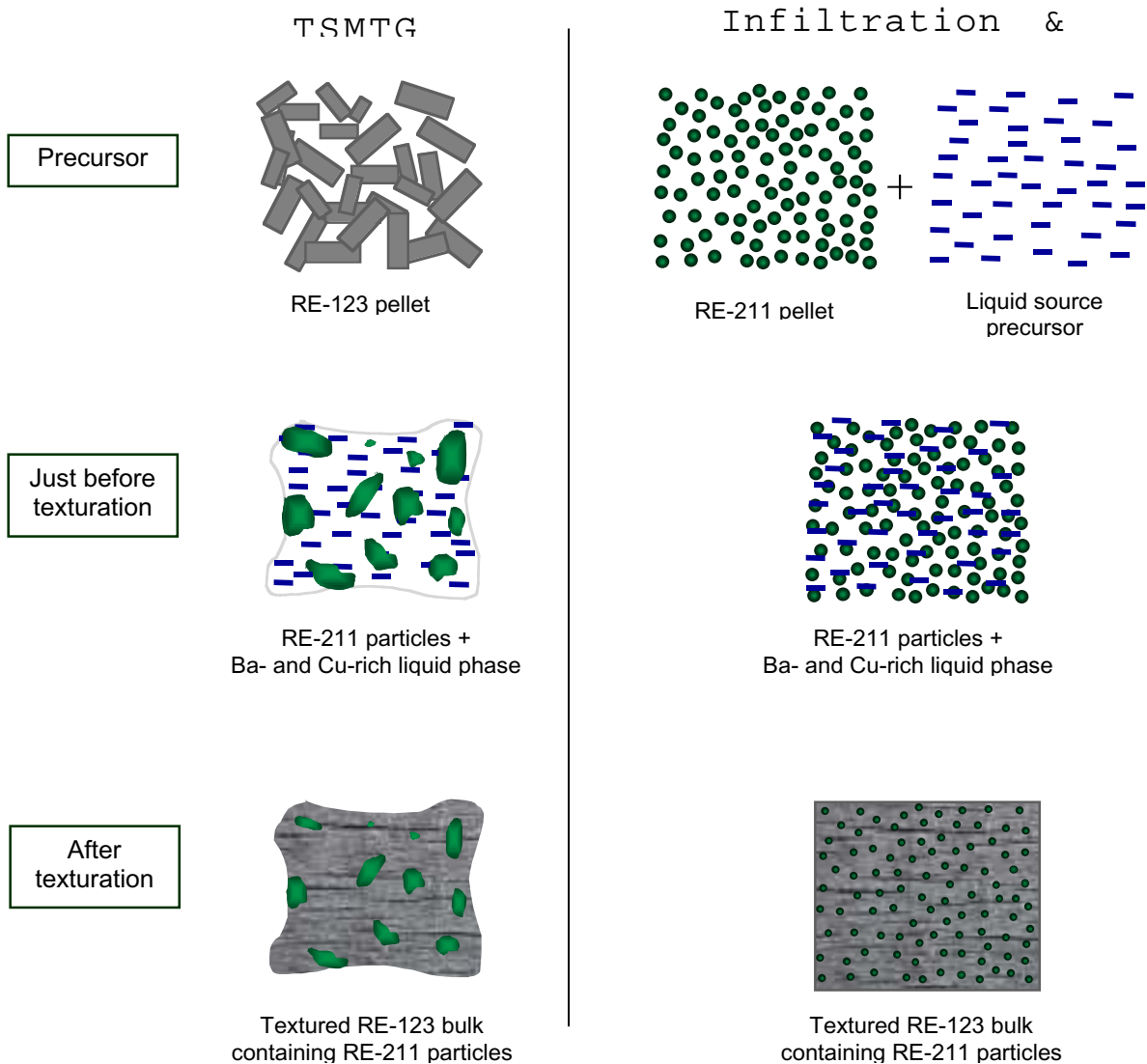

# Figure 19

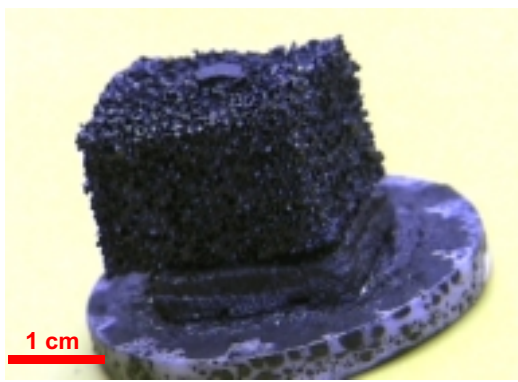